\documentstyle[12pt,dina4p,epsfig]{article}
\newcommand{\lwig}{\mbox{\,\raisebox{.3ex}
{$<$}$\!\!\!\!\!$\raisebox{-.9ex}{$\sim$}\,}}

\begin{document}
\begin{titlepage}
\title{{\normalsize{\rightline{DESY 94-193}\rightline{hep-ph/9411241}}}
\bigskip Infrared Fixed Points and Fixed Lines in the\\
       Top-Bottom-Tau Sector in\\Supersymmetric Grand Unification
\thanks{project supported in part by Deutsche Forschungsgemeinschaft}
       \vspace{7mm}}
\author{Barbara Schrempp\vspace{14mm}\\
        Universit\"at Kiel, Institut f\"ur Theoretische Physik\\
        Olshausenstr. 40, D-24098 Kiel\vspace{5mm}\\
        and \vspace{5mm}\\
        Deutsches Elektronen-Synchrotron DESY\\
        Notkestr.85, D-22603 Hamburg}
\date{October 1994}
\maketitle
\begin{abstract}
The two-loop ``top-down'' renormalization group flow for the top,
bottom and tau Yukawa couplings, from $\mu=M_{{\rm GUT}}\simeq
O(10^{16}\:{\rm GeV})$ to $ \mu\simeq m_{t}$, is explored in the
framework of supersymmetric grand unification; reproduction of the
physical bottom and tau masses is required. Instead of following the
recent trend of implementing {\it exact} Yukawa coupling unification
i) a search for infrared (IR) fixed lines and fixed points in the
$m_{t}^{{\rm pole}}\,$-$\,\tan\beta$ plane is performed and ii) the
extent to which these imply {\it approximate} Yukawa unification is
determined.  In the $m_{t}^{{\rm pole}}\,$-$\,\tan\beta$ plane two IR
fixed lines, intersecting in an IR fixed point, are located. The more
attractive fixed line has a branch of almost constant top mass,
$m_{t}^{{\rm pole}}\simeq 168$-$180\,{\rm GeV}$ (close to the
experimental value), for the large interval $2.5\lwig\tan\beta\lwig
55$; it realizes tau-bottom Yukawa unification at $M_{{\rm
GUT}}$ approximately. The less attractive fixed line as well as
the fixed point at $m_{t}^{{\rm pole}}\simeq 170\,{\rm GeV}$,
$\tan\beta\simeq 55$ implement approximate top-bottom Yukawa
unification at all scales $\mu$. The renormalization group flow is
attracted towards the IR fixed point by way of the more attractive IR
fixed line. The fixed point and lines are distinct from the much
quoted {\it effective} IR fixed point $m_{t}^{{\rm pole}}\simeq O(200
\,{\rm GeV})\sin\beta$.
\end{abstract}

\thispagestyle{empty}
\end{titlepage}
\newpage

There has been renewed interest in supersymmetric grand unification
triggered by the recent sucessful reevaluation of gauge coupling
unification \cite{ell}.  Also the appealing concepts of tau-bottom
Yukawa coupling unification, e.g. valid \cite{cha} in a minimal
$SU(5)$ theory, or possibly even of tau-bottom-top Yukawa unification
have been revived \cite{bar1}-\cite{lan1}. The main issue has been to
determine how {\it exact} tau-bottom (-top) Yukawa coupling
unification at the unification scale $\mu=M_{{\rm GUT}}\simeq
O(10^{16}\,{\rm GeV})$ constrains the low energy parameters of the
top-bottom-tau sector at the infrared (IR) scale $\mu=m_{t}$. In this
paper a reversed strategy is followed. First the unconstrained
``top-down'' renormalization group (RG) flow in the top-bottom-tau
sector, from $\mu=M_{{\rm GUT}}$ to $\mu=m_{t}$, is explored. The aim
is, to determine the IR fixed lines and IR fixed points in the low
energy top-bottom-tau sector, which attract the RG flow. Then the
extent to which these distinguished IR lines and points reflect {\it
approximate} Yukawa coupling unification at $M_{{\rm GUT}}$ is
investigated.

The simplest framework is the minimal supersymmetric extension of the
Standard Model (MSSM), embedded in a ``one-step'' grand unified theory
(e.g. an $SU(5)$ theory). It holds from $M_{{\rm GUT}}$ down to a
scale $\mu=M_{{\rm SUSY}}$ of the order of the weak interaction scale,
parametrizing effectively the thresholds of the supersymmetric
particles and the heavy Higgs bosons. For simplicity $M_{{\rm
SUSY}}=m_{t}$ is chosen in the following. The top-bottom-tau sector is
described by the ``top down'' RG evolution of the Yukawa couplings
$h_{t}(\mu)$, $h_{b}(\mu)$, $h_{\tau}(\mu)$, from $\mu=M_{{\rm GUT}}$
to $\mu=m_{t}$. The running top, bottom and tau masses at $\mu=m_{t}$ are
\begin{eqnarray}
m_{t}(m_{t})&=&h_{t}(m_{t})\,\sin\beta\, v/\sqrt{2},\label{mt}\\
m_{b}(m_{t})&=&h_{b}(m_{t})\,\cos\beta\, v/\sqrt{2},\label{mb}\\
m_{\tau}(m_{t})&=&h_{\tau}(m_{t})\,\cos\beta\, v/\sqrt{2}\label{mtau}.
\end{eqnarray}
Here $\tan\beta=v_{2}/v_{1}$ parametrizes the unknown ratio of the
vacuum expectation values $v_{1}$, $v_{2}$ of the neutral components
of the two Higgs doublets, with $v_{1}^2+v_{2}^2=v^2\simeq (246\, {\rm
  GeV})^2$. The physical top mass is given by the pole mass \cite{tar}
\begin{equation}
m_{t}^{{\rm pole}}=m_{t}(m_{t})(1 + \frac{4}{3 \pi}
\alpha_{3}(m_{t})+O(\alpha_3^2)).
\label{mtphys}
\end{equation}

In a ``top-down'' RG analysis of the top-bottom-tau sector there are
four free parameters to start with, $\tan\beta$ and e.g. $h_{t0}$,
$h_{b0}$, $h_{\tau 0}$, the ultraviolet (UV) initial values of the
Yukawa couplings at $\mu=M_{{\rm GUT}}$. Two IR conditions are
provided by the physical masses $m_{\tau}$, $m_{b}$ (evolved to
$\mu=m_{t}$ within the Standard Model). This leaves e.g. $\tan\beta$
and the IR parameter $m_{t}^{{\rm pole}}$ free.

Let me briefly recapitulate the consequences \cite{bar1}-\cite{lan1}
of implementing exact tau-bottom Yukawa coupling unification, $h_{\tau
0}=h_{b0}$, at $M_{{\rm GUT}}$. The allowed $\tan\beta$,
$m_{t}^{{\rm pole}}$ values are constrained to a narrow band in the
$m_{t}^{{\rm pole}}\,$-$\,\tan\beta$ plane, which turns out
\cite{bar1}-\cite{lan1} surprisingly to coincide essentially with the
triviality bound determined by the IR image of {\it large} UV initial
values $h_{t0}$, the much discussed strongly attractive IR fixed
point\footnote{Usually also Refs. \cite{pen,zim,sch1} are quoted along
with Ref. \cite{hil} in this context; however, it will be important in
the following that they refer to a different though nearby fixed point.}
\cite{hil}-\cite{all}
\begin{equation}
    m_{t}^{{\rm pole}}\simeq O(200\, {\rm GeV})\,\sin\beta.
\label{Hillfp}
\end{equation}
Compared with the direct \cite{cdf} and indirect \cite{lep}
experimental evidences for the top mass
\begin{equation}
  {\rm direct\, evidence}\;\;\;m_{t}=174\pm
  10\begin{array}{l}+13\\-12\\ \end{array}{\rm GeV},\;{\rm indirect\,
    evidence}\;\;\;m_{t}=173\begin{array}{l}+12+18\\-13-20\\
\end{array}{\rm GeV},
\label{CDFLEP}
\end{equation}
most likely only two narrow windows, one at small $\tan\beta\simeq O(1.5)$ and
one at large $\tan\beta\simeq O(60)$ are admitted. Top-bottom-tau
Yukawa unification is realized near $\tan\beta\simeq 60$.

For the following it is important to realize that the fixed point
(\ref{Hillfp}) is rather what has been called
\cite{hil} an {\it effective} fixed point, since neither does its position
remain fixed under the ``top down'' evolution nor does it attract RG
solutions from below.

In this paper the two-loop RG flow in $h_{t}(\mu)$, $h_{b}(\mu)$,
$h_{\tau}(\mu)$ for the ``top down'' evolution, from $M_{{\rm GUT}}$
to $m_{t}$, is investigated. This implies studying solutions starting
their evolution at {\it arbitrary} initial values $h_{t0}$, $h_{b0}$,
$h_{\tau 0}$ within the perturbatively allowed range $ h_{t0}^{2}/ 4
\pi$, $h_{b0}^{2}/ 4 \pi$, $h_{\tau 0}^{2}/ 4\pi <  O(1)$ and for
an {\it arbitrary} value of $\tan\beta$, only subject to the condition that
the physical masses $m_{b}$ and $m_{\tau}$ be reproduced. No Yukawa
coupling unification is imposed; so at $\mu=m_{t}$
essentially the whole $m_{t}^{{\rm pole}}\,$-$\,
\tan\beta$ plane, bounded from above by the triviality bound (the IR images
of large $h_{t0}$ or $h_{b0}$),
is available for investigation. The main issues are then
\begin{itemize}
\item to search for genuine IR fixed points and fixed lines in the
  $m_{t}^{{\rm pole}}\,$-$\,\tan\beta$ plane, which attract the RG
  flow from above as well as from below,
\item to determine the strength of their IR attraction and
  thus a hierarchy of their importance,
\item to investigate to which extent Yukawa coupling unification can be
  maintained on an approximate level.
\end{itemize}

Let me briefly anticipate the most interesting resulting IR fixed
manifolds in the $m_{t}^{{\rm pole}}\,$-$\,\tan\beta$ plane, the IR
fixed line $m_{t}^{{\rm pole}}\simeq (170$-$180)\,{\rm GEV}
\sin\beta$ as well as the IR fixed point $m_{t}^{{\rm
pole}}\simeq 170\,{\rm GeV}$, $\tan\beta\simeq 55\,$, both implying an
almost constant top mass $m_{t}^{{\rm pole}}\simeq 168$-$180\,{\rm GeV}$ for
the large interval $2.5\lwig\tan\beta\lwig 55$. This top mass is in
good agreement with the experimental top mass value
(\ref{CDFLEP}). The issue of top-bottom Yukawa coupling unification
will enter the discussion automatically, since one IR fixed line as
well as the IR fixed point imply it approximately for all scales
$\mu$. Tau-bottom Yukawa unification at $M_{{\rm GUT}}$ will turn out
to hold only approximately.

Let me mention in this context that IR fixed lines are of particular
interest, if they are more strongly attractive than IR fixed points.
This happens e.g. in the Standard Model, where an IR fixed line,
corresponding to a top-Higgs mass relation, has turned out \cite{sch1}
to be much more strongly attractive than the corresponding IR fixed
point \cite{pen,zim,sch1}. An extended analysis of the
Higgs-top-bottom-tau sector in the Standard Model is in preparation
\cite{sch2}.

Let me point out that what physicists like to call a ``fixed
line'' is termed in mathematical language \cite{gug} more appropriately
an ``invariant line''; this terminology has been used in Ref. \cite{sch1}.

The two-loop RG equations \cite{ino,bar1} of the top-bottom-tau sector
of the MSSM involve $g_3$, $g_2$, $g_1$, the $\,SU(3)\times
SU(2)\times U(1)\,$ gauge couplings, and the Yukawa couplings $h_{t}$,
$h_{b}$, $h_{\tau}$ running as functions of $\, t=\ln (\mu / M_{{\rm
GUT}})$, assuming vanishing u, d, s, c, e, $ \mu $ Yukawa couplings.
Before turning to the numerical two-loop RG analysis, considerable
{\it analytical insight} into the IR fixed manifolds will be obtained
from the corresponding {\it one-loop} RG equations with {\it vanishing
electroweak gauge couplings}, $g_1,\: g_2=0$. In this reduced
framework, which is a good approximation except for the small region
in coupling space, where $g_1,\:g_2$ are not dominated by {\it any} of
the other couplings, the IR fixed points and fixed lines are {\it
exact}. They will appear in the variables
\begin{equation}
\rho_{t}=\frac{h_{t}^2}{g_3^2},\hspace{1cm}\rho_{b}=\frac{h_{b}^2}{g_3^2},
\hspace{1cm}\rho_{\tau}=\frac{h_{\tau}^2}{g_3^2},
\label{rtbtau}
\end{equation}
and not in $h_{t}$, $h_{b}$, $h_{\tau}$. By eliminating the variable
$t$ in favour of $g_3^2$ the top-bottom-tau sector in the reduced one
-loop framework is described by the following RG equations in terms of
the three variables (\ref{rtbtau})
\begin{eqnarray}
  -3 g_3^2\frac{{\rm d}\,\rho_{t}}{{\rm d}\, g_3^2}&=&\rho_{t}\,
  (6\rho_{t}+\rho_{b}-\frac{7}{3}),
\label{diffrho1}\\
-3 g_3^2\frac{{\rm d}\,\rho_{b}}{{\rm d}\, g_3^2}&=&\rho_{b}\,
(6\rho_{b}+\rho_{t}+\rho_{\tau}-\frac{7}{3}),
\label{diffrho2}\\
-3 g_3^2\frac{{\rm d}\,\rho_{\tau}}{{\rm d}\, g_3^2}&=&\rho_{\tau}\,
(3\rho_{b}+4\rho_{\tau}+3).
\label{diffrho3}
\end{eqnarray}
Clearly, this system of coupled differential equations has four
(finite) fixed points, of which only the fourth one (\ref{fptbtau}) is
IR attractive
\begin{equation}
  \rho_{t}=0\;\, ({\rm IR\:
    repulsive}),\;\;\;\rho_{b}=0\;\, ({\rm IR\:
    repulsive}),\;\;\;\rho_{\tau}=0\;\, ({\rm IR\:
    attractive}),
\end{equation}
\begin{equation}
  \rho_{t}=\frac{7}{18}\;\, ({\rm IR\: attractive}),\;\;\;
  \rho_{b}=0\;\, ({\rm IR\:
    repulsive}),\;\;\;\rho_{\tau}=0\;\, ({\rm IR\:
    attractive}),
\label{fprt}
\end{equation}
\begin{equation}
  \rho_{t}=0\;\, ({\rm IR
    \:repulsive}),\;\;\;\rho_{b}=\frac{7}{18}\;\, ({\rm
    IR\: attractive})\;\;\;\rho_{\tau}=0\;\, ({\rm IR\:
    attractive}),
\label{fprb}
\end{equation}
\begin{equation}
  \rho_{t}=\frac{1}{3}\;\, ({\rm IR\:
    attractive}),\;\;\;\rho_{b}=\frac{1}{3}\;\, ({\rm IR\:
    attractive}),\;\;\;\rho_{\tau}=0\;\, ({\rm IR\:
    attractive}).
\label{fptbtau}
\end{equation}
The fixed point (\ref{fprt}) is the supersymmetric counterpart of a
well-known fixed point in the Standard Model \cite{pen,zim,sch1}. It is,
however,
much more IR attractive with respect to $\rho_{t}$, since it is
approached as
\begin{equation}
  \rho_{t}(g_3^2)=\frac{7/18}{\displaystyle
    1-(1-\frac{7}{18\rho_{t0}}) ( \frac{g_3^2}{g_{30}^2})^{-7/9}}
\label{-7/91}
\end{equation}
for $\rho_{b,\tau}=0$. The measure of the IR attraction is the size of
the negative power in $(g_3^2/g_{30}^2)^{-7/9}$, i.e. -7/9, to be compared
with the corresponding one, -1/7, in the Standard Model. The
discussion of the fixed point (\ref{fprb}) runs analogously, with the
roles of $\rho_{t}$ and $\rho_{b}$ interchanged.

Linearization of the system of differential equations
(\ref{diffrho1})-(\ref{diffrho3}) in the neighbourhood of the genuine
IR fixed point (\ref{fptbtau}) leads to the following approximate
analytical solution
\begin{eqnarray}
  \rho_{t}(g_3^2)&\simeq &\frac{1}{3}+\frac{1}{2}\bigg(
  (\rho_{t0}+\rho_{b0}-\frac{2}{3}-\frac{1}{5}\rho_{\tau 0})
  (\frac{g_3^2}{g_{30}^2})^{-7/9}+(\rho_{t0}-\rho_{b0}+\frac{1}{7}\rho_{\tau
    0}) (\frac{g_3^2}{g_{30}^2})^{-5/9}\nonumber\\ &
  &+\frac{2}{35}\rho_{\tau 0}
  (\frac{g_3^2}{g_{30}^2})^{-4/3}\bigg),\label{-7/92}\\
  \rho_{b}(g_3^2)&\simeq &\frac{1}{3}+\frac{1}{2}\bigg(
  (\rho_{t0}+\rho_{b0}-\frac{2}{3}-\frac{1}{5}\rho_{\tau 0})
  (\frac{g_3^2}{g_{30}^2})^{-7/9}-(\rho_{t0}-\rho_{b0}+\frac{1}{7}\rho_{\tau
    0}) (\frac{g_3^2}{g_{30}^2})^{-5/9}\nonumber\\ &
  &+\frac{12}{35}\rho_{\tau 0}
  (\frac{g_3^2}{g_{30}^2})^{-4/3}\bigg),\label{-7/93}\\
  \rho_{\tau}(g_3^2)&\simeq &\rho_{\tau 0}\,
  (\frac{g_3^2}{g_{30}^2})^{-4/3}\label{-4/3},
\end{eqnarray}
where $\rho_{t0}$, $\rho_{b0}$, $\rho_{\tau 0}$, $g_{3 0}^2$ are the
corresponding initial values at the UV scale.

Mathematical theorems \cite{gug} about systems of coupled differential
equations with several fixed points imply that the system
(\ref{diffrho1})-(\ref{diffrho3}) has {\it an IR attractive fixed
  plane (invariant plane)}, the $\rho_{t}$-$\rho_{b}$ plane,
$\rho_{\tau}(g_3^2)\equiv 0$, and in this plane {\it two IR fixed
  lines (invariant lines)} intersecting in the IR fixed point
(\ref{fptbtau}), $\rho_{t}=\rho_{b}=1/3$; the two fixed lines are
\begin{itemize}
\item $\rho_{t}(g_3^2)\equiv\rho_{b}(g_3^2)$, equivalent to
  $h_{t}(g_3^2)\equiv h_{b}(g_3^2)$, {\it implying exact top-bottom Yukawa
  coupling unification at all scales $\mu$},
\item the solution of Eqs. (\ref{diffrho1}), (\ref{diffrho2}) for
  $\rho_{\tau}=0$ which interpolates the three fixed points
  (\ref{fprt})-(\ref{fptbtau}), i.e. ($\rho_{t}=7/18,\:\rho_{b}=0$),
  ($\rho_{t}=1/3,\:\rho_{b}=1/3$), ($\rho_{t}=0,\:\rho_{b}=7/18$),
  having roughly the shape of two adjacent sides of a square.
\end{itemize}

The first fixed line can be read off from
\begin{equation}
  -3 g_3^2\frac{{\rm d}\, (\frac{\displaystyle \rho_{b}}{\displaystyle
      \rho_{t}})}{{\rm d}\, g_3^2}=5 \rho_{b}
  (\frac{\rho_{b}}{\rho_{t}}-1),
\end{equation}
as obtained from Eqs. (\ref{diffrho1}), (\ref{diffrho2}) for
$\rho_{\tau}=0$. As may be inferred from Eqs.
(\ref{-7/91})-(\ref{-4/3}), the fixed $\rho_{t}\,$-$\,\rho_{b}$ plane,
$\rho_{\tau}\equiv 0$, is IR attractive like $(g_3^2/g_{30}^2)^{-4/3}$
near the fixed point (\ref{fptbtau}), the square-type fixed line is
the more strongly attractive one, attracting like $(g_3^2 /
g_{30}^2)^{-7/9}$ in its neighbourhood, the $\rho_{t}=\rho_{b}$ fixed
line is the less attractive one, attracting like $(g_3^2 /
g_{30}^2)^{-5/9}$ near the IR fixed point (\ref{fptbtau}).

Altogether, the ``top-down'' RG flow is attracted towards the IR fixed
point (\ref{fptbtau}) by way of first being attracted towards the
square-type IR fixed line in the $\rho_{t}$-$\rho_{b}$ plane and then
(practically) along this line towards the fixed point. Of course,
solutions with initial values on an IR fixed line are drawn along it
towards the IR fixed point.

Reinstating the electroweak couplings $g_1,\:g_2$ and using the full
two-loop RG equations \cite{ino,bar1}, the fixed points and fixed
lines are found not to be exact any more. In order to determine the
{\it approximate} positions of the fixed points and fixed lines one
could follow Ref. \cite{pen} and replace $g_1,\:g_2$ by constants,
given by their averages along the evolution path. This introduces a
weak dependence on the evolution path, in particular on the UV scale.
Here, it is preferred to maintain properly running couplings
$g_1(\mu),\:g_2(\mu)$ and to follow a prescription, which has already
been successfully applied in Ref. \cite{sch1}. For a given evolution
path, e.g. from $\mu= M_{{\rm GUT}}$ to $\mu= m_{t}$,
\begin{itemize}
\item the approximate IR fixed point in the $\rho_{t}$-$\rho_{b}$ plane
 is determined as the unique point which has {\it the same
    value} at $\mu= M_{{\rm GUT}}$ and at $\mu= m_{t}$,
\item an approximate IR fixed line in the $\rho_{t}$-$\rho_{b}$ plane
  is determined by the condition that all points, which {\it start on
    it} or - in case of the most attractive line -  {\it sufficiently close
    to it} at $\mu= M_{{\rm GUT}}$, {\it end on it} at $\mu= m_{t}$.
\end{itemize}
Again the results depend somewhat on the evolution path, in particular on
the UV scale; the numerical deviations from the positions given in
Eqs. (\ref{fprt})-(\ref{fptbtau}) are the larger, the larger the UV
scale.

Obviously, the scenario of grand unification is not really vital
for the results. The IR fixed lines and fixed point are basically IR
properties of the MSSM. Only to the extent that they are not exact but
approximate do they depend on the evolution path, i.e. on the size of
$M_{{\rm GUT}}$.

The following numerical RG analysis uses for $\mu\leq m_{t}$ the same
Standard Model evolutions for $g_1,\,g_2,\,g_3,\,m_{\tau},\,m_{b}$ as
Ref. \cite{bar1}, comprising three-loop expressions for $\alpha_3$ and
$m_{b}$, with slightly different actualized input parameters.  For
$\mu\geq m_{t}$ the two-loop RG evolution equation of the MSSM
\cite{ino,bar1} are used for vanishing first and second generation
Yukawa couplings. No error analysis is performed.  The inputs are as
follows
\begin{itemize}
\item The supersymmetry scale is chosen at
  $M_{{\rm SUSY}}=m_{t}=174\,{\rm GeV}$.
\item From Ref. \cite{par} \,$m_{Z}=91.187\,{\rm GeV},\;$ $\sin^2
  \hat{\theta}_{Z}(m_{Z})=0.2319,\;$ $\alpha_{{\rm em}}(m_{Z})=127.9$
  leading to $g_1^2(m_{Z})=0.2132,\;$ $g_2^2(m_{Z})=0.4237$.
\item The grand unification scale $M_{{\rm GUT}}$ as well as
  $\alpha_3(m_{Z})$ are determined by requiring unification of all three
  gauge couplings at $\mu=M_{{\rm GUT}}$ (neglecting the two-loop
  Yukawa coupling contributions in the gauge coupling RG equations,
  following Ref. \cite{bar1}). The results are $M_{{\rm GUT}}=2.223 \,10^{16}
  \,{\rm GeV}$, $g_1^2=g_2^2=g_3^2=0.5278$ at $M_{{\rm GUT}}$ and
  $\alpha_3(m_{Z})=0.1247$.
\item
\begin{equation}
  m_{\tau}=1.7771\,{\rm GeV}\;\cite{par}\;\;\;{\rm and}\;\;\;m_{b}=
  4.25\,{\rm GeV}\;\cite{gas},
\label{mbmtau}
\end{equation}
are used, admitting a variation within $4.1\,{\rm
GeV}\leq m_{b}\leq 4.4\,{\rm GeV}$ \cite{bar1}. The evolutions to
$\mu=m_{t}$ are evaluated to be given by
\begin{equation}
  m_{b}(m_{t})=m_{b}/\eta_{b},\;\; m_{\tau}
  (m_{t})=m_{\tau}/\eta_{\tau}\;\;{\rm with}\;\;\eta_{b}=1.599,
  \;\;\; \eta_{\tau}=1.02.
\label{masses}
\end{equation}
\item RG evolutions are performed from $\mu=M_{{\rm GUT}}$ to
$\mu=m_{t}=174\,{\rm GeV}$.
\end{itemize}

Of primary interest is to investigate the RG flow in the variables
$\rho_{t}=h_{t}^2/g_3^2$, $\rho_{b}=h_{b}^2/g_3^2$, $\rho_{\tau}=
h_{\tau}^2/g_3^2$ from $M_{{\rm GUT}}$ to $m_{t}$ (subject to the
condition that the physical tau and bottom masses be reproduced), to
locate the IR fixed point and fixed lines in the $\rho_{t}$-$\rho_{b}$
plane, and to display their respective strenghts of attraction.

This is achieved in Fig.\,1 as follows: 36 pairs of initial values
($\rho_{t0},\,\rho_{b0}$) are chosen along the boundary of a square, defined
by its four corners (0.0001,\,0.0001), (0.0001,\,25), (25,\,25),
(25,\,0.0001); the value 25 is the limit of validity of perturbation
theory. The evolution from $\mu=M_{{\rm GUT}}$ to $\mu=m_{t}$ maps
the 36 initial UV points onto 36 final IR points, which are the
starting points of 36 polygons in Fig.\,1. The 36 IR points are taken
as a new set of UV initial values, evolved to a new set of IR final
points. This procedure is repeated until all final points meet in the
IR fixed point. This leads to 36 sequences of evolutions. In order to
guide the eye, the IR points of each sequence are connected by
straight lines to form the 36 polygons displayed in Fig.\,1.

For {\it each} of these evolutions the corresponding pair of initial
values ($\rho_{t0}$, $\rho_{b0}$) is complemented by a $\rho_{\tau 0}$
value, which is adjusted such that after the evolution
the appropriate ratio $\rho_{\tau}(m_{t})/\rho_{b}(m_{t})=m_{\tau}^2
(m_{t})/m_{b}^2 (m_{t})$ at $\mu= m_{t}$ is reproduced, which is fixed
by Eqs. (\ref{mbmtau}) and(\ref{masses}). This requires values for
$\sqrt{\rho_{\tau 0}/\rho_{b0}}=h_{\tau 0}/h_{b0}$ ranging between
0.92 and 2.1.

The paths which the polygons take in the $\rho_{t}\,$-$\,\rho_{b}$
plane in Fig. 1 demonstrate that there is a rather strongly IR
attractive fixed line. It has two branches, labelled by ``1'' and
``2'' in Fig. 1. Branch ``1'' has almost constant $\rho_{t}$ and is of
physical interest; it runs from $\rho_{t}=0.689$, $\rho_{b}=0$ (the
analogon of the fixed point (\ref{fprt}) and the supersymmetric
counterpart of the fixed point in Refs. \cite{pen,zim,sch1}) to the IR
fixed point at $\rho_{t}=.609$, $\rho_{b}=.505$; branch ``2'', being
almost constant in $\rho_{b}$, runs from $\rho_{t}=0$,
$\rho_{b}=0.586$ (the analogon of the fixed point (\ref{fprb})) to the
fixed point. A less attractice IR fixed line, labelled by ``3'' in
Fig. 1, is seen to be $\rho_{t}\simeq 1.2\,\rho_{b}$, intersecting
the more attractive IR fixed line in the IR fixed point. This line
corresponds to $h_{t}\simeq 1.1\,h_{b}$, implying approximate
top-bottom Yukawa coupling unification at all scales $\mu$. The IR
fixed point as well as the two IR fixed lines are as expected from the
preceding analytical considerations only with shifted positions due to
the effect of including the electroweak gauge couplings and to a
lesser extent to the two-loop contributions. The lengths of the
sections of the polygons are a direct measure of the strength of IR
attraction: the longer the section, the stronger the attraction.
Clearly, as expected, the RG flow is primarily attracted towards the
more attractive IR fixed line and then runs (practically) along it
towards the IR fixed point. The IR final points coming from UV initial
points with $\rho_{t0}=25$ or $\rho_{b0}=25$ are connected by a dashed
line; this line represents the triviality bound.  In the neighbourhood
of branch ``1'' as well as near the IR fixed point, a ratio
$\sqrt{\rho_{\tau 0}/\rho_{b0}} =h_{\tau 0}/h_{b0}\simeq 1.3 $ is
required, which realizes tau-bottom Yukawa coupling unification at
$M_{{\rm GUT}}$ only approximately.

Fig.\,1 in terms of the variables $\rho_{t}=h_{t}^2/g_3^2$ and
$\rho_{b}=h_{b}^2/g_3^2$ is translated\footnote{by translating the IR
points of Fig.\,1 and then connecting them again by polygons} into
Fig.\,2 for the variables $m_{t}^{{\rm pole}}$ and $\tan\beta$ by
means of Eqs. (\ref{mt}), (\ref{mb}), (\ref{mtphys}), (\ref{mbmtau})
and (\ref{masses}) and by enforcing $m_{b}=4.25\,{\rm GeV}$. The maps
of the infrared fixed lines are again labelled consistently by ``1'',
``2'' and ``3''. The dotted line is again the triviality bound, one
branch of which is the effective IR fixed point (\ref{Hillfp}).

The more strongly attractive IR fixed line, branch ``1'', is not
sufficiently attractive to draw all perturbatively allowed initial
values $h_{t0} < 3.6$ onto it, given an RG evolution from $M_{{\rm
GUT}}$ to $m_{t}$. However, it is remarkable that all initial values
$0.6\lwig h_{t0} < 3.6$ are drawn {\it from above and from below}
within a strip corresponding to $m_{t}^{{\rm pole}}\simeq 174\pm
23\,{\rm GeV}\sin\beta$. The IR fixed point is less attractive.

Fig 3 shows the dependence of the results on $m_{b}$: for the values
$m_{b}=4.1,\, 4.25,\, 4.4\,{\rm GeV}$ the pairs of IR fixed lines
intersecting in their common IR fixed points $m_{t}^{{\rm
pole}}=169.6\,{\rm GeV}$, $\tan\beta=57.4,55.4,53.5$, respectively,
are displayed. Clearly, the physically interesting branch ``1'' is
hardly affected by this variation.

The rich IR fixed manifolds, displayed in Figs. 1 and 2, are
summarized according to their manifestations in the $h_{t}$-$h_{b}$
plane resp. in the $m_{t}^{{\rm pole}}\,$-$\,\tan\beta$ plane
\begin{itemize}
\item a more strongly attractive IR fixed line (invariant line) with
  the two branches
\begin{itemize}
\item[i)] branch ``1'':
\begin{equation}
  h_{t}\simeq .93-.99\;\;\;{\rm
resp.}\;\;\;m_{t}^{{\rm pole}}\simeq (170-180)\,{\rm
GeV}\sin\beta,
\label{175}
\end{equation}
implying an almost constant top mass $m_{t}^{{\rm pole}}\simeq 168-180\,{\rm
GeV}$ for the large interval $2.5\lwig\tan\beta\lwig 55$; in its neighborhood
approximate tau-bottom Yukawa coupling unification at $\mu=M_{{\rm GUT}}$ is
realized
\begin{equation}
h_{\tau 0}\simeq 1.3\,h_{b0}.
\label{viol2}
\end{equation}
\item[ii)] branch ``2'':
\begin{equation}
  h_{b}\simeq .85- .91\;\;\;{\rm
resp.}\;\;\;\tan\beta\simeq 55-60,
\label{58}
\end{equation}
\end{itemize}
\item a less strongly attractive IR fixed line (invariant line),
denoted by ``3'',
\begin{equation}
  h_{t}\simeq 1.1\, h_{b}\;\;\;\;{\rm resp.}\;\;\;\;m_{t}^{{\rm
pole}}\simeq 3\,{{\rm GeV}}\tan\beta,
\label{uni}
\end{equation}
implying approximate top-bottom Yukawa coupling unification at all
scales $\mu$,
\item an attractive IR fixed point at their intersection
\begin{equation}
h_{t}\simeq 0.93,\;h_{b}\simeq 0.85\;\;\;\;{\rm resp.}\;\;\;\;
m_{t}^{{\rm pole}}\simeq 170\,{\rm GeV},\;\tan\beta\simeq 55,
\label{170}
\end{equation}
implying both, approximate top-bottom Yukawa unification at all scales
$\mu$, $h_{t}\simeq 1.1\,h_{b}$, as well as approximate tau bottom
Yukawa unification, $h_{\tau 0}\simeq 1.3\,h_{b0}$, at $\mu=M_{{\rm
GUT}}$.
\end{itemize}
The ``top-down'' RG flow is attracted towards the IR fixed point by
way of the more attractive IR fixed line.

 From this list a physically appealing solution can be extracted with
the following properties:
\begin{itemize}
\item it is part of the most attractive IR fixed line ending in the IR
fixed point, which both attract the ``top-down'' RG flow
from above and from below,
\item it corresponds to an almost constant top mass value,
$m_{t}^{{\rm pole}}\simeq 168$-$180\,{\rm GeV}$, matching the
experimental top mass,
\item it holds for the large range $2.5\lwig\tan\beta\lwig 55$,
\item it implies approximate tau-bottom Yukawa unification at
$M_{{\rm GUT}}$, at the level $h_{\tau 0}\simeq 1.3\,h_{b0}$,
\item for $\tan\beta\simeq 55$ it fulfills in addition approximate
top-bottom Yukawa unification at all scales $\mu$, at the level
$h_{t}\simeq 1.1\,h_{b}$.
\end{itemize}
Let me emphasize that the constant top mass value of 168-180
GeV as well as the tau-bottom (-top) Yukawa coupling
unification, even though valid only approximately, are encoded in the
positions of the IR fixed manifolds (\ref{175})-(\ref{170}) and are
not put in by hand at any stage.

The solution described above is rather dominated by IR aspects. It has
to be contrasted with the somewhat complementary solution
\cite{bar1}-\cite{lan1} discussed in the introduction which is rather
dominated by UV aspects : implementation of exact tau-bottom Yukawa
unification,$h_{\tau 0}=h_{b0}$, at $M_{{\rm GUT}}$, which drives the
low energy solution towards the triviality bound, requiring large
initial UV values $h_{t0}$, and admitting only two $\tan\beta$
windows, $\tan\beta\simeq O(1.5)$ and $\tan\beta\simeq O(60)$.

It would be interesting to investigate whether threshold effects and
the influence of higher dimensional operators at $M_{{\rm GUT}}$,
which have been considered \cite{lan2} as possible sources for slight
violations of exact gauge coupling unification, could be responsible
for violations of exact Yukawa coupling unification to the extent
implied in Eqs. (\ref{viol2}) and (\ref{uni}).

Of course all numerical results depend to some extent on the input
parameters chosen, in particular on the position of $M_{{\rm SUSY}}$
and the value of $\alpha_3(m_{Z})$. A more extended analysis, varying
these parameters, is in progress; as expected, no major changes in the
resulting IR fixed structures in the $m_{t}^{{\rm pole}}$-$\tan\beta$
plane are found.

\vspace{2mm}
\noindent
{\bf Acknowledgements:} I thank F. Schrempp for constructive comments
after reading the ma\-nu\-script. I also thank V. Barger and M.S.
Berger for a helpful communication. I am grateful to the DESY Theory
Group for the extended hospitality.

\begin{figure}
\begin{center}
\unitlength1mm
\begin{picture}(140,140)
\epsfig{file=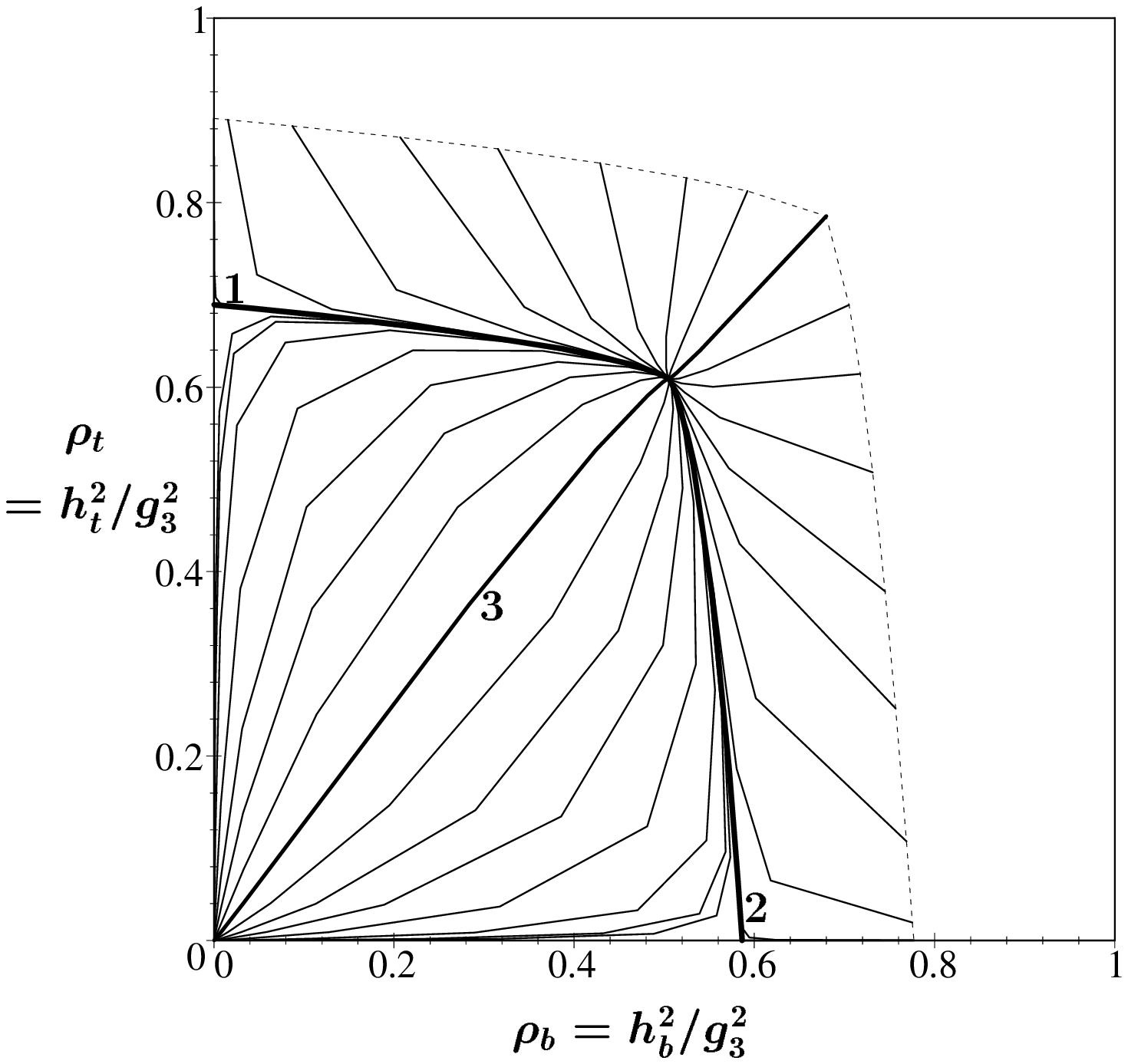,bbllx=76pt,bblly=354pt,bburx=495pt,bbury=751pt,height=14cm,width=14cm}
\end{picture}
\caption[rtrb]{The two-loop renormalization group flow from $\mu=M_{GUT}$
to $\mu=174\,{\rm GeV}$ in the $\rho _{t}$ -$\rho _{b}$ plane: first
towards the more IR attractive fixed line with the two branches,
labelled by ``1'' and ``2'', and then practically along it towards the
IR fixed point at $\rho _{t}= .609$, $\rho_{b}= .505$.  The less IR
attractive fixed line, $\rho_{t}\simeq 1.2\,\rho_{b}$, labelled by
``3'', as well as the IR fixed point automatically imply approximate
top-bottom Yukawa coupling unification. The dashed curve is the
triviality bound.}
\end{center}
\end{figure}

\begin{figure}
\begin{center}
  \unitlength1mm
\begin{picture}(140,140)
\epsfig{file=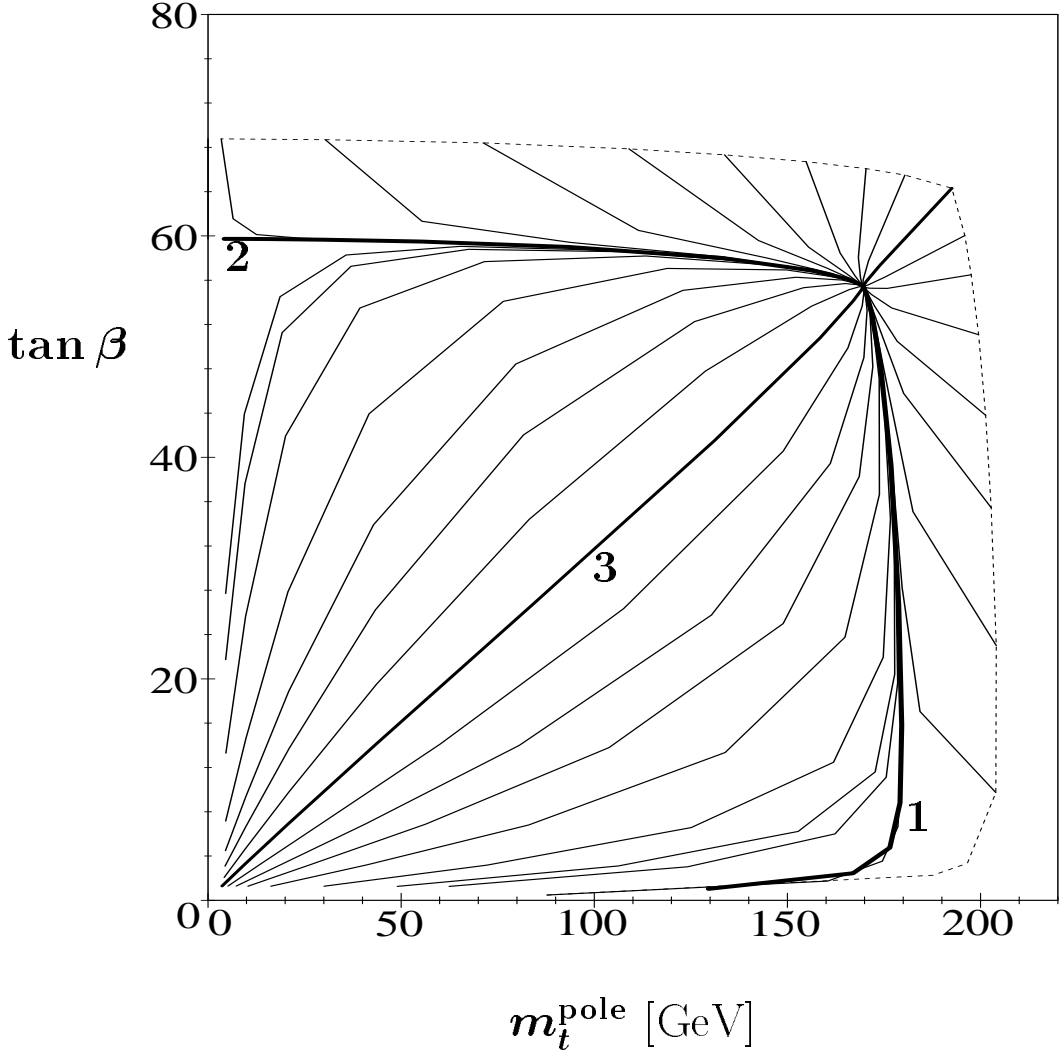,bbllx=90pt,bblly=221pt,bburx=464pt,bbury=617pt,height=14cm,width=14cm}
\end{picture}
\caption[mtphys]{The two-loop renormalization group flow from $\mu=M_{GUT}$
to $\mu=174\,{\rm GeV}$ in the $\tan\beta$-$m{_t}^{{\rm pole}}$ plane,
as obtained from Fig.\,1 for $m_{b}=4.25\,{\rm GeV}$\,: first towards
the IR attractive fixed line, with the two branches $m_{t}^{{\rm
pole}}\simeq (170$-$180)\,{\rm GeV}\sin\beta$\, labelled by ``1'', and
$\tan\beta\simeq 0.55-0.60$, labelled by ``2'', and then practically
along it towards the IR fixed point at $m_{t}^{{\rm pole}}=170\,{\rm
GeV}$, $\tan\beta=55$.  Of physical interest is the branch ``1''
including the fixed point, which imply an almost constant top mass
$m_{t}^{{\rm pole}}\simeq 168$-$180\,{\rm GeV}$ for
$2.5\lwig\tan\beta\lwig 55$; along this branch approximate tau-bottom
Yukawa coupling unification is realized at $M_{{\rm GUT}}$. The less
attractive IR fixed line $m_{t}^{{\rm pole}}\simeq 3\,{\rm
GeV}\,\tan\beta$, labelled by ``3'', as well as the IR fixed point
automatically imply approximate top-bottom Yukawa coupling unification
at all scales $\mu$.  The dashed curve is the triviality bound; one of
its branches corresponds to the well-known {\it effective} IR fixed
point $m_{t}^{{\rm pole}}=O(200\,{\rm GeV})\sin\beta$.}
\end{center}
\end{figure}

\begin{figure}
\begin{center}
  \unitlength1mm
\begin{picture}(140,140)
\epsfig{file=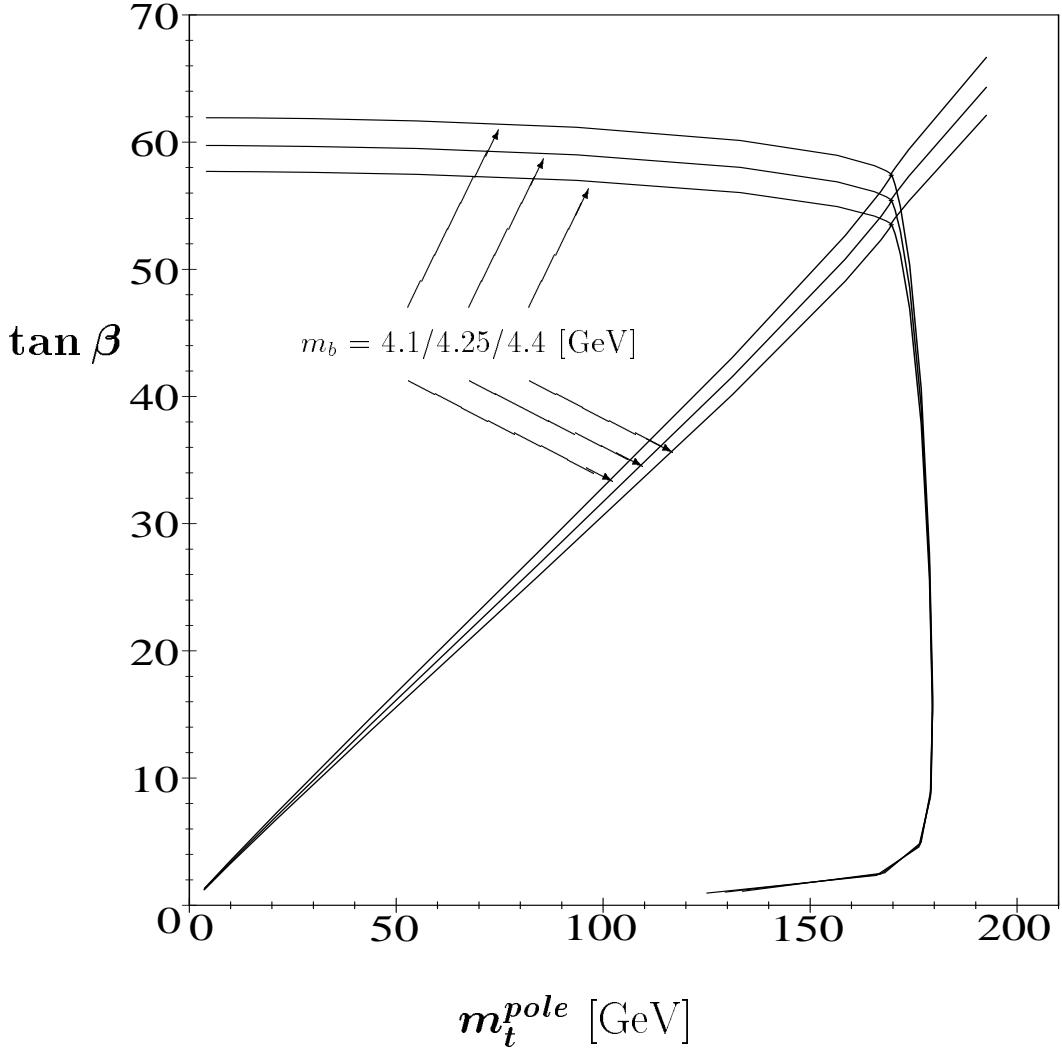,bbllx=98pt,bblly=220pt,bburx=495pt,bbury=624pt,height=14cm,width=14cm}
\end{picture}
\caption[invli]{The more IR attractive fixed lines (square type) and the less
IR attractive fixed lines (straight line type) for $m_{b}=4.1,\, 4.25,\,
4.4\,{\rm GeV}$, intersecting in their respective IR fixed points
$\tan\beta=57.4,\, 55.4,\, 53.5$, $m_{
t}^{{\rm pole}}=170\,{\rm GeV}$. The less IR attractive fixed line as well as
the IR fixed point automatically imply approximate top-bottom Yukawa coupling
unification.}
\end{center}
\end{figure}

\end{document}